\documentclass[aps,prl,twocolumn]{revtex4}

\usepackage{graphicx, xcolor}
\usepackage{float}

\newcommand{\jl}[1]{\textcolor{black}{#1}}

\begin{document}

\title{Piecewise omnigenous stellarators}

\author{J.L. Velasco}

\affiliation{Laboratorio Nacional de Fusi\'on, CIEMAT, 28040 Madrid, Spain}

\author{I. Calvo}
\affiliation{Laboratorio Nacional de Fusi\'on, CIEMAT, 28040 Madrid, Spain}

\author{F.I. Parra}
\affiliation{Princeton Plasma Physics Laboratory, Princeton, NJ 08540, USA}

\author{F.J. Escoto}
\affiliation{Laboratorio Nacional de Fusi\'on, CIEMAT, 28040 Madrid, Spain}

\author{E. S\'anchez}
\affiliation{Laboratorio Nacional de Fusi\'on, CIEMAT, 28040 Madrid, Spain}

\author{H. Thienpondt}
\affiliation{Laboratorio Nacional de Fusi\'on, CIEMAT, 28040 Madrid, Spain}

\date{\today}

\begin{abstract}

In omnigeneous magnetic fields, charged particles are perfectly confined in the absence of collisions and turbulence. For this reason, the magnetic configuration is optimized to be close to omnigenity in any candidate for a stellarator fusion reactor. However, approaching omnigenity imposes severe constraints on the spatial variation of the magnetic field. In particular, the topology of the contours of constant magnetic-field-strength on each magnetic surface must be such that there are no particles transitioning between different types of wells. This, in turn, usually leads to complicated plasma shapes and coils. This Letter presents a new family of optimized fields that display tokamak-like collisional energy transport while having transitioning particles. This result radically broadens the space of accessible reactor-relevant configurations.

\end{abstract}

\maketitle

\begin{figure}[h!]
\vskip-0.2cm\includegraphics[angle=0,width=0.47\columnwidth]{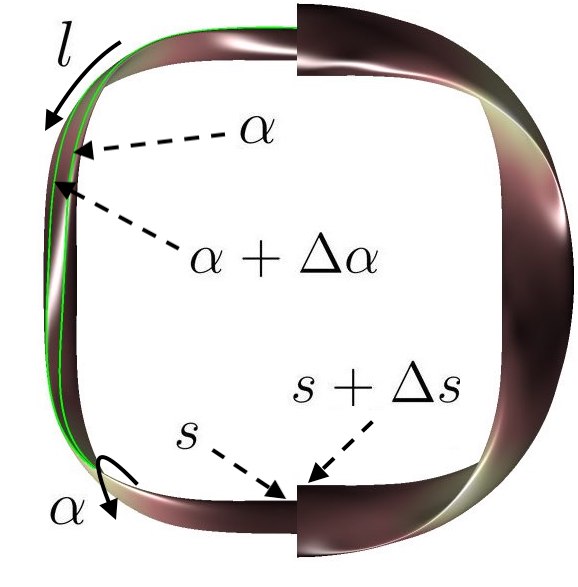}~~~
\includegraphics[angle=0,width=0.49\columnwidth]{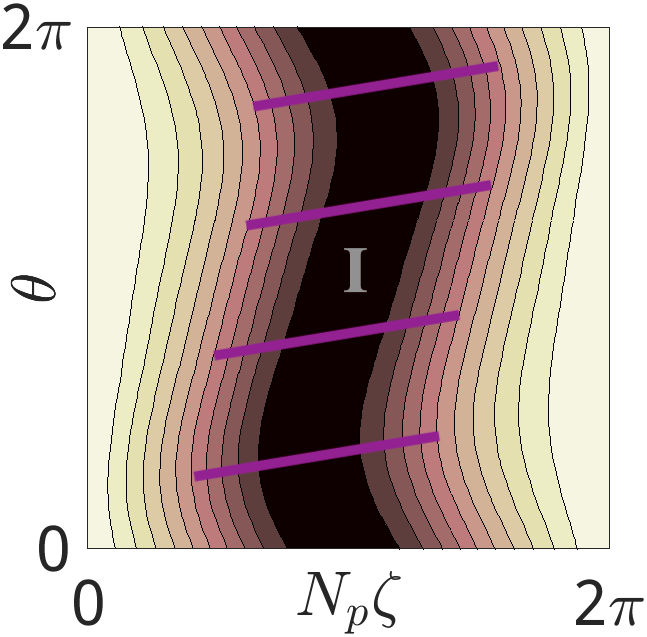}
\includegraphics[angle=0,width=0.49\columnwidth]{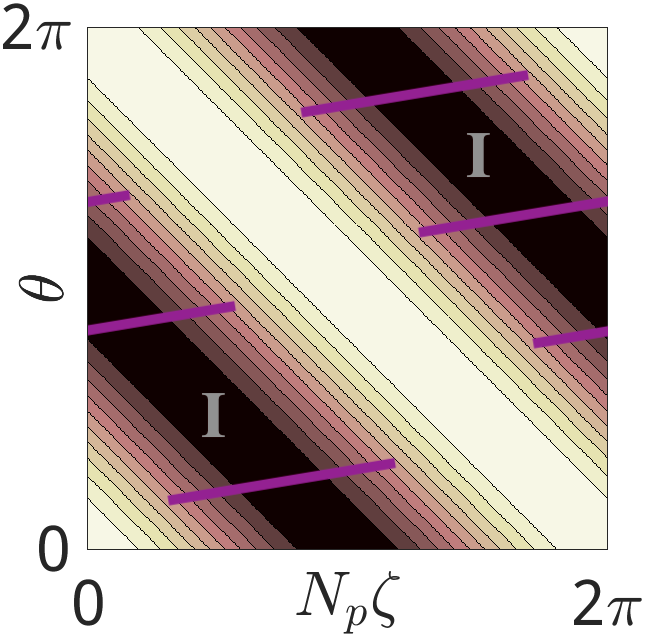}
\includegraphics[angle=0,width=0.49\columnwidth]{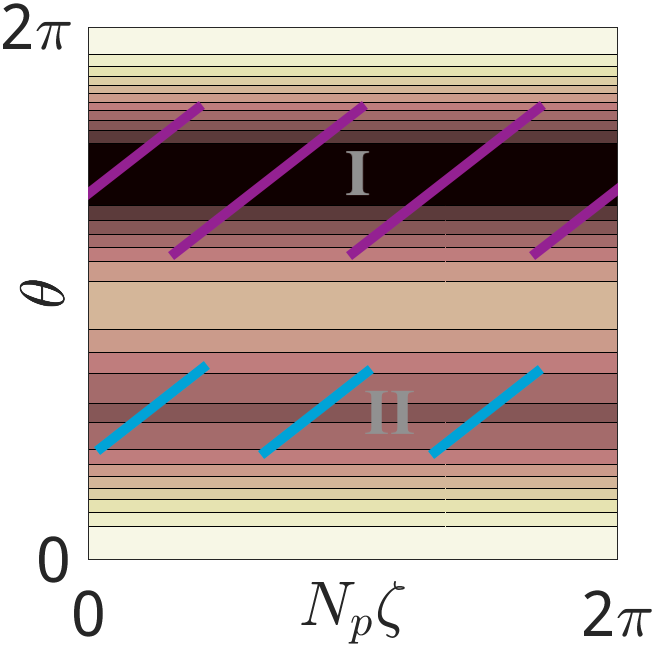}
\caption{Top view of two flux surfaces of a stellarator magnetic configuration, with \jl{two} magnetic field lines on the same flux surface highlighted \jl{in green} (top left). Contours of constant $B$ for several omnigenous fields: quasi-isodynamic (top right), quasi-helically symmetric (QHS) with $(N,M)=(1,-1)$ (bottom left), quasi-axisymmetric with two contours of maximum $B$ (bottom right). A darker colour corresponds to a weaker field, and roman numerals label regions with different classes of orbits (sketched in purple and cyan, parallel to $\mathbf{B}$).\label{FIG_SKETCH_OMNI}}
\end{figure}

Stellarators confine charged fusion reactants and products in a three-dimensional toroidal geometry consisting of nested magnetic surfaces (figure~\ref{FIG_SKETCH_OMNI}, top left), also called \textit{flux surfaces}, in which the magnetic field $\mathbf{B}$ is generated by external magnets. This makes stellarators easier to operate than tokamaks, in which an important part of the magnetic field is created by an inductive plasma current. This current \jl{can drive} disruptive instabilities and complicates steady-state operation, which may limit the viability of the tokamak concept in the path towards reliable and efficient commercial fusion power plants.

The three-dimensionality of the stellarator magnetic field \jl{(which lacks the axisymmetry of the tokamak)} comes at the price of generally having worse plasma confinement. Both in tokamaks and stellarators, in the absence of collisions and turbulence, charged passing particles (i.e., those whose component of the velocity that is parallel to $\mathbf{B}$ never vanishes) go over the complete flux surface and are, on average, tied to it. Conversely, charged trapped particles \textit{live} in regions of low magnetic field strength $B$, moving back and forth along a magnetic field line (a motion parametrized by the arc length $l$) at constant energy $\mathcal{E}$ and magnetic moment $\mu$. On a longer time scale, they drift perpendicularly to $\mathbf{B}$. This motion takes place in the radial direction, i.e. perpendicularly to the flux surface, and/or tangentially to it. A field is omnigenous \cite{cary1997omni} if, for all trapped particles, the drift in the radial direction averages to zero. In such a field, charged particles are perfectly confined in the absence of collisions. Axisymmetric tokamaks are omnigenous, which explains their small collisional loss of energy and particles, known as neoclassical transport. Stellarators can be \textit{neoclassically optimized} to be nearly omnigenous, so that the very deleterious, stellarator-specific, $1/\nu$ neoclassical transport remains low~\cite{beidler2021nature}.

Approaching omnigenity requires a very careful tailoring of the magnetic configuration, as it imposes stringent constraints on the spatial variation of $B$ on the flux surface, illustrated in figure~\ref{FIG_SKETCH_OMNI}. Cary and Shasharina~\cite{cary1997omni} derived a list of such constraints, and a thorough discussion can also be found e.g.~in \cite{landreman2012omni}. First, all contours of constant $B$ on the flux surface must close in the toroidal, poloidal or helical direction. Moreover, the distance along the field line between the bounce points (the points where ${\mathcal{E}}/\mu=B$ and thus the component of the velocity that is parallel to $\mathbf{B}$ becomes zero) cannot depend on the field line. A consequence derived from these constraints is that the contour of maximum $B$, $B_{\mathrm{max}}$, must be a straight line when the flux surface is parametrized in \jl{Boozer poloidal and toroidal angles, $\theta$ and $\zeta$ (these coordinates map each flux surface in a way that field lines are straight lines with a slope given by the rotational transform $\iota$). Specifically,  $B=B_{\mathrm{max}}$ along a line} of constant $M\theta-N_pN\zeta$, with $M$ and $N$ integers, and $N_p$ the number of field periods \jl{(such that $B(\theta,\zeta)=B(\theta,\zeta+2\pi/N_p)$)}. Omnigenous fields with $M=0$ are called quasi-isodynamic (QI) (figure~\ref{FIG_SKETCH_OMNI}, top right). Quasisymmetric (QS) fields (figure~\ref{FIG_SKETCH_OMNI}, bottom left) are omnigenous fields in which all $B$-contours are straight lines in Boozer coordinates. The vast majority of the most recently designed stellarator magnetic configurations, e.g.~\cite{henneberg2019qa,plunk2019direct,kinoshita2019cfqs,bader2020wistell,landreman2022preciseQS,landreman2022mapping,camachomata2022direct,jorge2022qi,sanchez2023qi,goodman2023qi,dudt2023omni}, are approximately omnigenous fields that have been obtained by following explicitly some formulation of the constraints described by~\cite{cary1997omni}. Finally, less constrained omnigenous fields have been discussed \cite{parra2015omni}, with more than one contour of local minima and maxima, all of them still closed toroidally (figure~\ref{FIG_SKETCH_OMNI}, bottom right), poloidally or helically. This feature is key in order not to have \textit{transitioning particles} \cite{cary1997omni,parra2015omni} (an example \jl{of transitioning particles} would be particles drifting between regions I and II in figure~\ref{FIG_SKETCH_OMNI}, bottom right), which may have a deleterious effect on transport.

\begin{figure}
\includegraphics[angle=0,width=0.47\columnwidth]{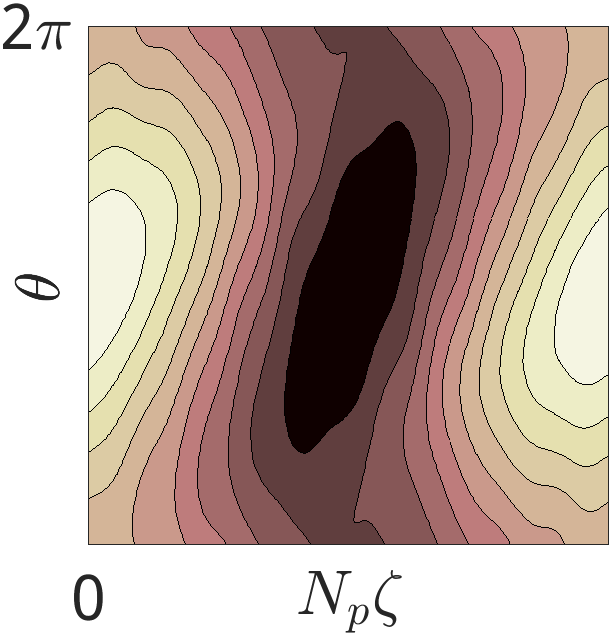}~
\includegraphics[angle=0,width=0.48\columnwidth]{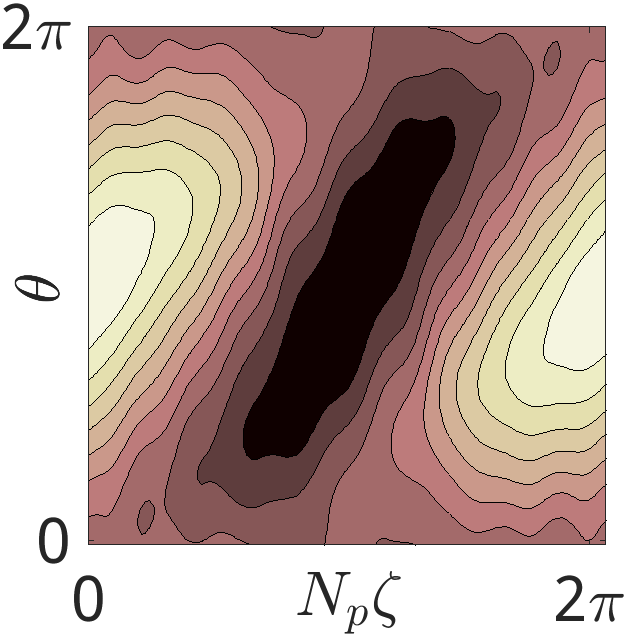}
\includegraphics[angle=0,width=0.49\columnwidth]{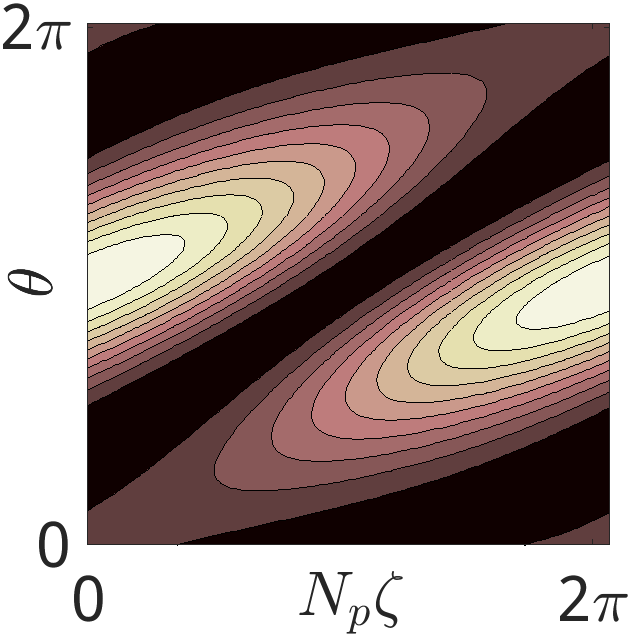}
\includegraphics[angle=0,width=0.49\columnwidth]{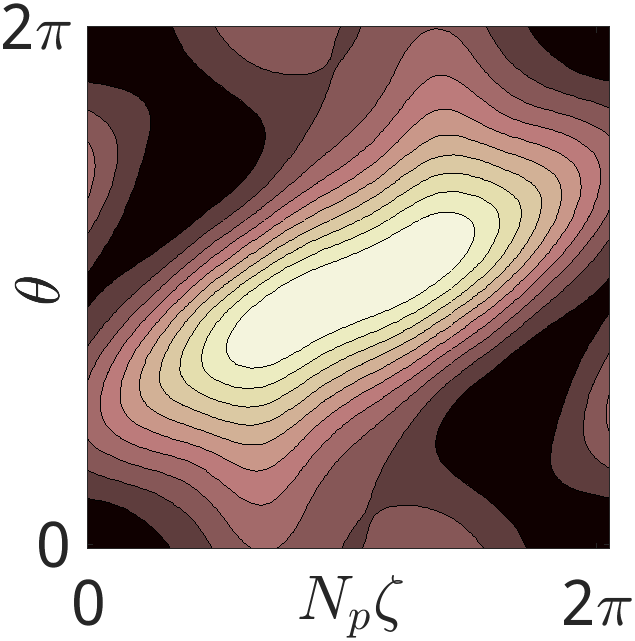}
\caption{$B$ on flux surface $s=0.25$ of the high mirror (top left) and standard (top right) configurations of W7-X, LHD inward-shifted (bottom left) and configuration A of \cite{bindel2023direct}.\label{FIG_W7XLHD}}
\end{figure}

\smallskip

An appropriate radial variation of $B$ helps retain good transport properties even at a distance from omnigenity, see e.g.~\cite{velasco2023flatmirror}. Still, it is of great interest to relax even further these constraints, which contribute to complicate the construction and operation of a fusion device. For example, the complexity of the coils needed to produce a nearly QS field was the main reason for the cancellation of the NCSX project \cite{strykowsky2009ncsx}. Fortunately, there are indications that relaxing these requirements is actually feasible. \jl{For instance, the stellarator with best experimental energy confinement (measured by the configuration-dependent factor in the ISS04 confinement scaling)} is LHD \cite{yamada2005taue}, whose neoclassically-optimized configuration (called inward-shifted, figure \ref{FIG_W7XLHD}, bottom left) has $B$-contours that do not have the topology discussed in~\cite{cary1997omni,parra2015omni}. \jl{Furthermore,} the coils of the stellarator Wendelstein 7-X (W7-X) (and of its extrapolations to a reactor \cite{beidler2001hsr4}) have been designed to produce an approximately QI field (high-mirror configuration, figure \ref{FIG_W7XLHD}, top left, to be compared to figure \ref{FIG_SKETCH_OMNI}, top right). \jl{However, there exists another magnetic configuration of W7-X which has better neoclassical transport while being further from exact quasi-isodynamicity (standard configuration, figure \ref{FIG_W7XLHD}, top right), as most $B$-contours do not close poloidally. This improved neoclassical transport, along with reduced turbulence, has been key for achieving the best performances of W7-X, with record values of the fusion triple product for a stellarator \cite{beidler2021nature}}.

Transport in these configurations, sometimes considered to be rough approximations to omnigenity, is often discussed in terms of a fraction of the trapped particles having approximately zero radial drift, see e.g.~\cite{mynick1982sigma,spong1998jstar}. In this Letter, by taking the analysis one step further, we identify and characterize a new family of magnetic fields with exceedingly small neoclassical transport that we call piecewise omnigenous (pwO) fields (defined rigorously in equations (\ref{EQ_PWO}) and (\ref{EQ_TRANSITIONS})). For pwO fields, the topology of the contours of constant magnetic-field-strength is different than for omnigenous fields  \cite{cary1997omni,parra2015omni}. This broadens significantly the space of neoclassically-optimized stellarator configurations and may lead to better reactors.

\smallskip

\begin{figure} 
\includegraphics[angle=0,width=0.49\columnwidth]{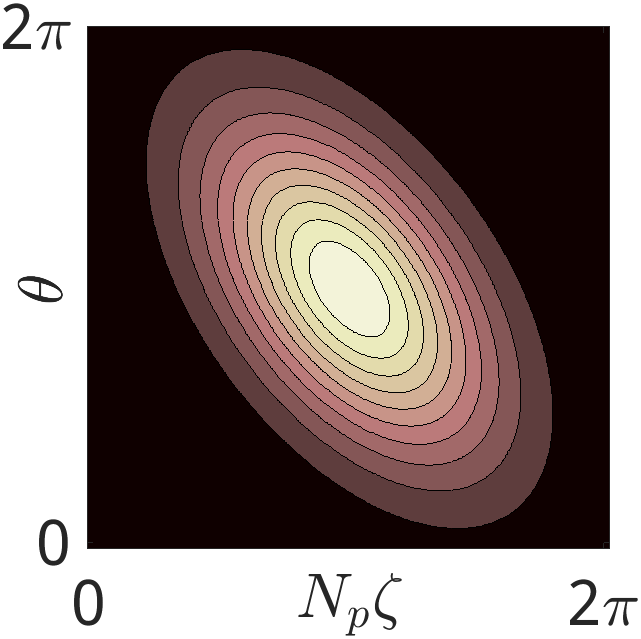}  
\includegraphics[angle=0,width=0.49\columnwidth]{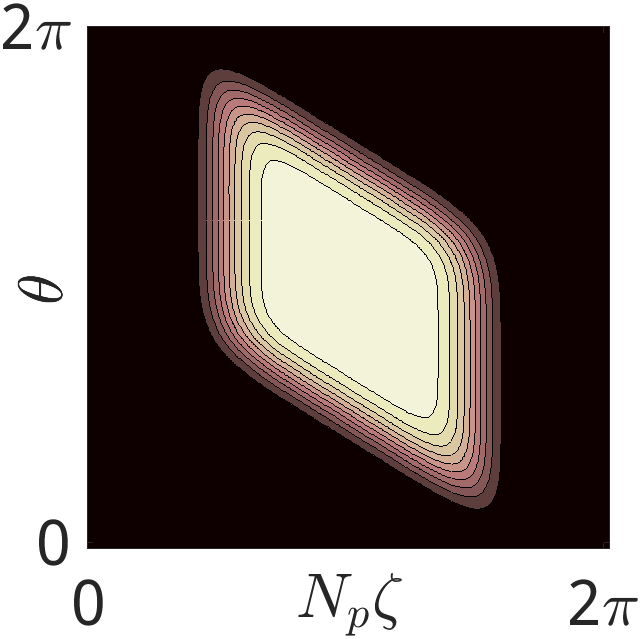}
\includegraphics[angle=0,width=1\columnwidth]{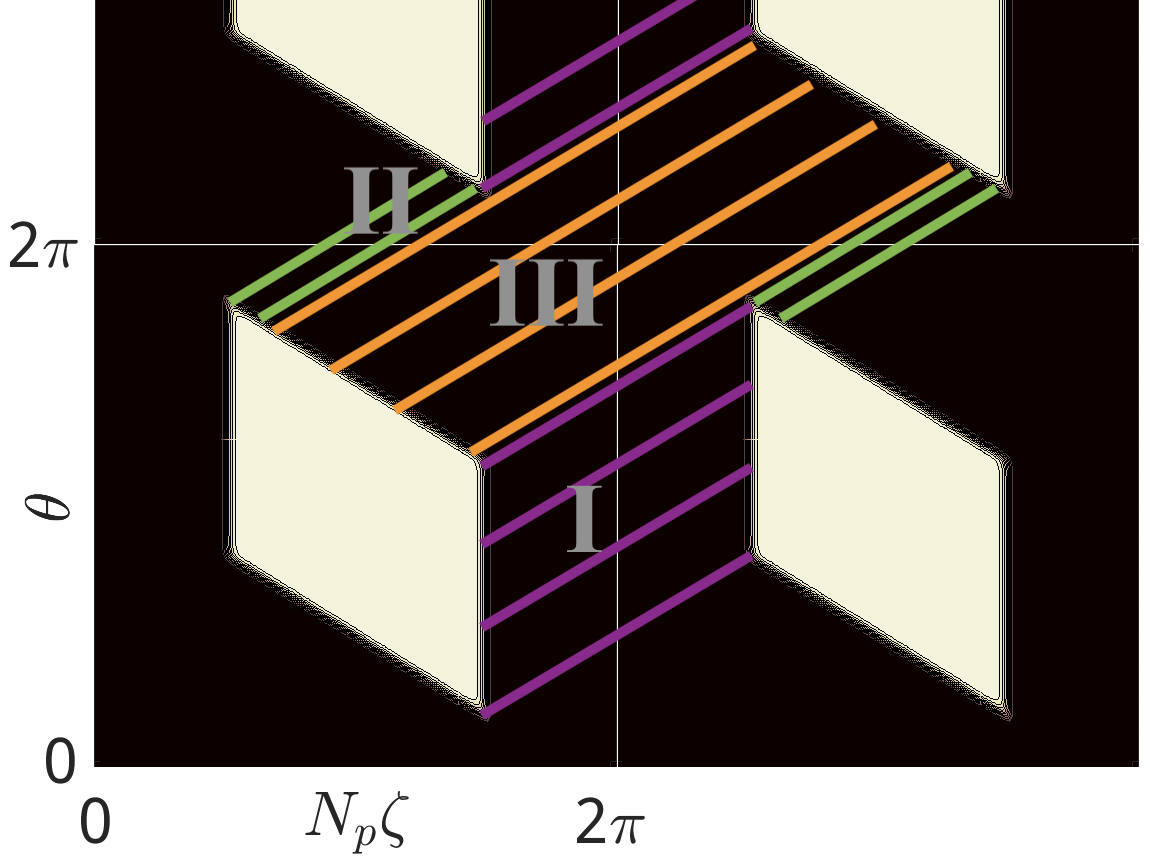}  
\caption{$B$ of equations (\ref{EQ_BPWO}) and (\ref{EQ_BPWOc}) with $N_pw_1=w_2=\pi/2$, $t_1=0$, $t_2=\iota_0=\iota=1.242$, $\zeta_c=\pi/N_p$ and $\theta_c=\pi$ for $p=1$ (top left), $p=3$ (top right) and $p=15$ (bottom).\label{FIG_PWO}}
\end{figure}

Figure \ref{FIG_W7XLHD} (bottom right) shows $B(\theta,\zeta)$ of configuration A of \cite{bindel2023direct}, recently obtained by direct optimization of guiding center orbits and showing very good neoclassical transport of both fast and bulk species, despite being visibly far from the omnigenous fields described in \cite{cary1997omni,parra2015omni}. It somewhat resembles LHD (compare with figure \ref{FIG_W7XLHD} bottom left shifted by $\pi$), but some relevant features are more marked for configuration A: the $B$-contours are, to some extent, composed of segments that are straight, but not of constant $M\theta-N_pN\zeta$. These contours are tightly distributed, i.e., $B$ is close to its minimum value, $B_{\mathrm{min}}$ or to $B_{\mathrm{max}}$ on a large fraction of the flux surface. Inspired by this, we construct a pwO field as the limit $p\rightarrow\infty$ of
\begin{eqnarray} 
\hskip-0.05cm
B=B_{\mathrm{min}}+\quad\quad\quad\quad\quad\quad\quad\quad\quad\quad\quad\quad\quad\quad\quad\quad\quad\quad\nonumber\\
(B_{\mathrm{max}}-B_{\mathrm{min}})e^{-\left(\frac{\zeta-\zeta_c+t_1(\theta-\theta_c)}{w_1}\right)^{2p}-\left(\frac{\theta-\theta_c+t_2(\zeta-\zeta_c)}{w_2}\right)^{2p}}\quad\hskip-0.15cm\label{EQ_BPWO}
\end{eqnarray}
in $0\le N_p\zeta< 2\pi$, $0\le\theta< 2\pi$ (to ensure continuity at $\zeta=0$ and $\theta=0$ and periodicity, $B$ from equation (\ref{EQ_BPWO}) is decomposed in its Fourier components, and a finite number of modes is kept). The field of equation (\ref{EQ_BPWO}) is pwO if the magnetic field lines follow a particular direction. This is set by the value of the rotational transform $\iota$, which we choose to be~\footnote{The cases $\iota=-t_2$ or $\iota=-1/t_1$, instead of equation (\ref{EQ_BPWOc}), are also of interest, and will be discussed elsewhere.}
\begin{eqnarray}
\iota=\iota_0\equiv \left(\frac{\pi(1-t_1t_2)}{N_pw_1}-1\right)^{-1}t_2\,.\label{EQ_BPWOc}
\end{eqnarray}
Figure~\ref{FIG_PWO} shows $B(\zeta,\theta)$ for one choice of the parameters $t_i$, $w_i$, $\zeta_c$ and $\theta_c$ (see caption; \jl{the particular role of each parameter is not relevant for the scope of this Letter, and will be discussed elsewhere \cite{velasco2024pwOb}}). \jl{The $B$-contours somewhat resemble those of figure \ref{FIG_W7XLHD} (bottom left) for small values of $p$.} As the limit $p\rightarrow\infty$ is approached, $B$ fulfills two conditions that will be sufficient (although not necessary \cite{velasco2024pwOb}) to achieve piecewise omnigenity:
\begin{enumerate}
\vskip-0.5cm\item All the contours $B_{\mathrm{min}}< B < B_{\mathrm{max}}$ collapse into a single parallelogram.\vspace{-0.2cm}
\item The rotational transform is such that only two field lines connect the four corners (which may be located in different field periods).
\end{enumerate}

Several relevant properties can be derived from these two conditions. In region I, the left bounce point ($\zeta_{b_1},\theta_{b_1}$) of an orbit (purple) lies on one of the sides of the $B$-contour if and only if the right bounce point ($\zeta_{b_2},\theta_{b_2}$) lies on the side parallel to it. Other regions exist on the flux surface, II and III, with different classes of trapped particles (green and orange, respectively). Within a given region, the distance between bounce points, $\zeta_{b_2}-\zeta_{b_1}$ is constant, and all particles see the same $B(\zeta-\zeta_{b_1})$ along their trajectory ($B$ has two values, $B_{\mathrm{max}}$ and $B_{\mathrm{min}}$).

For very large $p$, the field $B$ of equation (\ref{EQ_BPWO}) varies on a very small length scale $L$, and in the limit $p\rightarrow\infty$ it has discontinuities. Moreover, like the omnigenous fields of \cite{cary1997omni,parra2015omni}, it also has discontinous derivatives for $p\rightarrow\infty$. It must then be interpreted, like the Cary-Shasharina construction is, as an ideal design goal. For this reason, we first discuss omnigenity in the limit $p\rightarrow\infty$. Then, we demonstrate that the reduction of neoclassical transport already takes place for nearly piecewise omnigenous fields that are smooth, i.e., for fields of sufficiently large $p$, such that $L$ is small compared to the size of the flux surface, but sufficiently large compared to the Larmor radius, so that neoclasical theory can be rigorously applied.

\smallskip

Omnigenity can be best discussed in terms of the second adiabatic invariant $J(s,\alpha,{\mathcal{E}},\mu)$, defined as
\begin{equation}
J \equiv 2\int_{\zeta_{b_1}}^{\zeta_{b_2}}\mathrm{d}\zeta \frac{I_p+\iota I_t}{B^2} \sqrt{2\left(\mathcal{E}-\mu B\right)}\,,\label{EQ_J}
\end{equation}
with $I_t(s)=\mathbf{B}\cdot\mathbf{e}_\theta$ and $I_p(s)=\mathbf{B}\cdot\mathbf{e}_\zeta$. Here, $s$ is a radial coordinate proportional to the toroidal flux through the surface ($s=1$ at the last closed flux surface\jl{, at the plasma boundary}) and $\alpha=\theta-\iota\zeta$ \jl{is an angle that labels field lines on the flux surface}. The orbit-averaged radial drift is proportional to $\partial_\alpha J$, so having $\partial_\alpha J=0$ for all trapped particles is a sufficient condition for a magnetic field to be omnigenous. This is the case for all the points on the flux surfaces depicted in figure \ref{FIG_SKETCH_OMNI}, that follow the approach of~\cite{cary1997omni,parra2015omni}. \jl{In the deleterious $1/\nu$ regime, the source of the drift-kinetic equation, that describes collisional transport in stellarator plasmas, is proportional to $\partial_\alpha J$ and inversely proportional to the collision frequency. When $\partial_\alpha J = 0$, there is no $1/\nu$ transport and the stellarator, like the axisymmetric tokamak, displays the so-called banana regime, which sets a finite lower bound to neoclassical transport in toroidal devices \cite{landreman2012omni}}.

 In this work we broaden the discussion of omnigenity to fields for which $\partial_\alpha J=0$ is fulfilled \textit{piecewisely}, and for which there are regions of phase space at which $J$ has discontinuities in the $\alpha$ direction:
\begin{eqnarray}
J= J^\mathrm{(w)}(s,{\mathcal{E}},\mu),~~\mathrm{w}= \mathrm{I,II,III,...}\label{EQ_PWO}\\
\lim_{\Delta\alpha\to 0}J(s,\alpha+\Delta\alpha,{\mathcal{E}},\mu)-J(s,\alpha,{\mathcal{E}},\mu) \ne 0\nonumber \\ \mathrm{at~junctures.}\label{EQ_TRANSITIONS}
\end{eqnarray}
Equations (\ref{EQ_PWO}) and (\ref{EQ_TRANSITIONS}) constitute our definition of piecewise omnigenity, and the example field of figure~\ref{FIG_PWO} (bottom) follows them (with $\mathrm{w}=\mathrm{I,II,III}$ and $J^\mathrm{(I)}+J^\mathrm{(II)}=J^\mathrm{(III)}$). This definition relies on the existence of different regions of constant $J$, labelled with the discrete index $\mathrm{w}$ in equation (\ref{EQ_PWO}), and on the fact that trapped particles may transit between regions when precessing on the flux surface. These particles thus experience a discontinuity in $J$ at specific points of phase space that we will call \textit{junctures} \cite{dherbemont2022las}, as indicated by equation (\ref{EQ_TRANSITIONS}).

From equation (\ref{EQ_PWO}), it follows straightforwardly that $\partial_\alpha J=\partial_\alpha J^\mathrm{(w)}=0$ within a given region $\mathrm{w}$. Far from the juncture, trapped particles behave exactly as in an omnigenous field. The distinctive feature of piecewise omnigenous fields is that transitioning particles exist, but particles in the vicinity of the juncture still behave omnigenously. In a generic stellarator, in the vicinity of the juncture, the distance between bounce points depends on the field line (compare figure~\ref{FIG_PWO} (top right) and (bottom)) and this gives transport. Imposing equation (\ref{EQ_PWO}), thus making the $B$-contours have corners, guarantees that this is not the case in a piecewise omnigenous field. Finally, the discrete values of $\alpha$ at which $J$ is discontinuous \jl{do not give $1/\nu$ transport. There, the value of $\partial_\alpha J$ has to be interpreted considering all the orbits that converge to the juncture (see \cite{calvo2015flowdamping,nemov1999neo} for a detailed treatment of junctures in the $1/\nu$ regime of  generic stellarator fields). Since $(J^\mathrm{(I)}+J^\mathrm{(II)})-J^\mathrm{(III)}=0$, the source term of the drift-kinetic equation vanishes everywhere and there is no $1/\nu$ regime.}

\jl{We finally note that, although the choice of equation (\ref{EQ_BPWO}) is specially convenient for the discussion of this Letter, more general fields exist \cite{velasco2024pwOb} that comply with equations (\ref{EQ_PWO}) and (\ref{EQ_TRANSITIONS}). For instance, one could replace $B_\mathrm{min}$ with $B_\mathrm{I}(\zeta-t_1\theta)<B_\mathrm{max}$ in region I and $B_\mathrm{II}(\theta-t_2\zeta)<B_\mathrm{max}$ in regions II and III.}

\begin{figure} 
\includegraphics[angle=0,width=1.05\columnwidth]{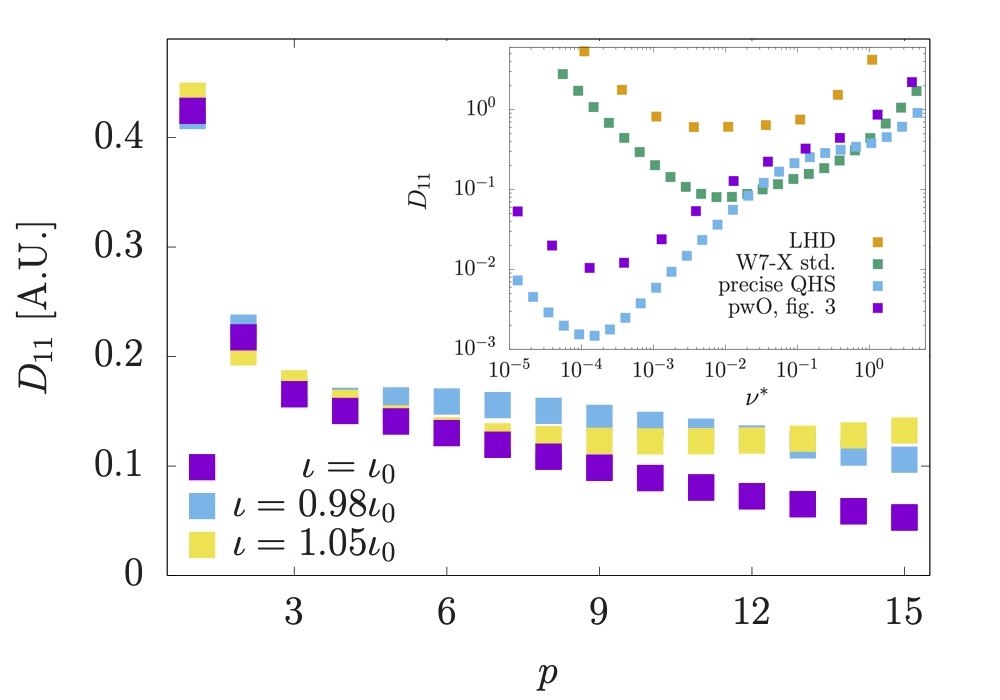}  
\caption{Neoclassical \jl{radial transport coefficient $D_{11}$} for the field of figure \ref{FIG_PWO} as a function of $p$ for $\iota=0.98\iota_0$, $\iota=\iota_0$ and $\iota=1.05\iota_0$. Inset: \jl{$D_{11}$} as a function of the collisionality $\nu^*$ (for the pwO case, $B_{\mathrm{max}}/B_{\mathrm{min}}-1=0.1$ and $N_p=2$).}\label{FIG_SCAN}
\end{figure}

\smallskip

We next discuss neoclassical transport in nearly pwO fields. Figure \ref{FIG_SCAN} shows that the monoenergetic \jl{$D_{11}$ transport coefficient (a standard figure of merit of radial neoclassical transport defined e.g.~in \cite{beidler2011ICNTS}, computed here with \texttt{MONKES}~\cite{escoto2024monkes})} decreases by roughly an order of magnitude from $p=1$ to $p=15$ for $\iota=\iota_0$ (and slighly less for an interval around $\iota_0$). This demonstrates that transport can be reduced in a controlled manner thanks to piecewise omnigenity, that the effect is significant already far from the ideal case, and in particular that it has some resilience against $\iota$ changes (recall the second condition for piecewise omnigenity). The inset of figure \ref{FIG_SCAN} shows a collisionality scan for the case $p=15$, $\iota=\iota_0$, for which transport at low-collisionality is roughly two orders of magnitude smaller than for the W7-X standard configuration \jl{and LHD}, and only one order of magnitude larger than for a \textit{precise} QHS field \cite{landreman2022preciseQS} (that deviates from that in figure \ref{FIG_SKETCH_OMNI} in $\sim$0.01\%, an unprecentedly small value). For a wide range of collisionalities, the neoclassical flux increases with $\nu^*$, as in a tokamak.

\begin{figure}
\includegraphics[angle=0,width=\columnwidth]{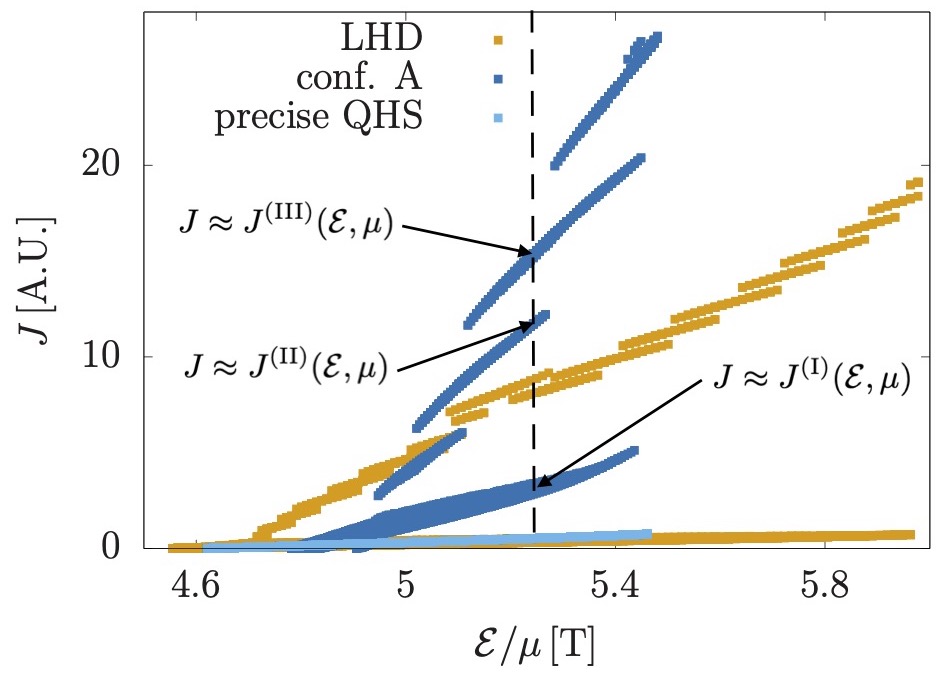}
\caption{$J$ as a function of $\mathcal{E}/\mu$ for $\sim 100$ field lines on surface $s=0.25$ of several configurations scaled to reactor size. \label{FIG_J}}
\end{figure}

\smallskip

For piecewise omnigenity to have a practical interest, it is additionally necessary to demonstrate that it can be achieved to a sufficient degree while being consistent with a reactor design (in particular, with being part of a \jl{magnetohydrodynamic} equilibrium created with feasible coils). In order to illustrate that this is the case, figure \ref{FIG_J} shows $J$ (computed with \texttt{KNOSOS}~\cite{velasco2021knosos}) for representative flux surfaces of configuration A of \cite{bindel2023direct}, LHD and the precise QHS. For the first two configurations (differently to the last one), for each velocity $\mathcal{E}/\mu$, $J$ clusters around three discrete values that fulfill $J^\mathrm{(I)}+J^\mathrm{(II)}\approx J^\mathrm{(III)}$. The other condition for piecewise omnigenity, the existence of transitioning particles, follows from the fact that most of the $B$-contours do not close toroidally, poloidally or helically, see figure \ref{FIG_W7XLHD} (bottom). LHD turns out to be a particularly simple nearly pwO configuration, with a large scale separation (in terms of $\zeta_{b_2}-\zeta_{b_1}$) between region I and regions II and III. 


\smallskip

\jl{Summarizing, in this Letter we have introduced the concept of piecewise omnigenity and, for the first time, provided a framework that explains how magnetic configurations with very small neoclassical transport such as W7-X standard and LHD inward-shifted, can exist very far from the usual description of omnigenity \cite{cary1997omni}. Even though pwO fields are optima of neoclassical transport, state-of-the-art optimization suites will not naturally find them, since they usually minimize targets (e.g.~the flux surface variance of $J$) that do not reach a mimimum value in a pwO field.}

\smallskip

\jl{The results presented in this Letter may radically expand the range of configurations that can be candidates for stellarator fusion reactors. Magnetic fields that are approximately pwO but not particularly close to the ideal pwO case show low neoclassical energy transport, something that does not usually happen in QI or QS fields~\cite{beidler2011ICNTS}, which may facilitate its construction in compliance with reactor requirements~\cite{miyazawa2014ffhrd1}. Moreover, as we discuss in some detail in the supplemental material, nearly pwO configurations could exhibit efficient alpha particle heating combined with low helium ash retention~\cite{bader2019fastions,bindel2023direct}. Finally, some nearly pwO configurations could display particular advantages. For instance, LHD has relatively low levels of ion-temperature-gradient (ITG) turbulence~\cite{regana2021imp3d}, a transport channel that has limited the performance of W7-X~\cite{bozhenkov2020hp}.}

\smallskip

\smallskip

\begin{acknowledgments}
This work has been carried out within the framework of the EUROfusion Consortium, funded by the European Union via the Euratom Research and Training Programme (Grant Agreement No 101052200 – EUROfusion). Views and opinions expressed are however those of the author(s) only and do not necessarily reflect those of the European Union or the European Commission. Neither the European Union nor the European Commission can be held responsible for them. This research was supported by grant PID2021-123175NB-I00, Ministerio de Ciencia, Innovaci\'on y Universidades, Spain. This work was supported by the U.S. Department of Energy under contract number DE-AC02-09CH11466. The United States Government retains a non-exclusive, paid-up, irrevocable, world-wide license to publish or reproduce the published form of this manuscript, or allow others to do so, for United States Government purposes. The authors are indebted to M. Padidar, M. Landreman and D. Bindel for providing configuration A, and had very helpful discussions with D. Spong, \jl{C. Zhu} and  EUROfusion's TSVV12  team. 
 \end{acknowledgments}

\vskip-0.35cm

\bibliography{/Users/velasco/Work/PAPERS/bibliography.bib}

\end{document}


\title{Supplemental material for\\ ``Piecewise omnigenous stellarators"}

\author{J.L. Velasco}

\affiliation{Laboratorio Nacional de Fusi\'on, CIEMAT, 28040 Madrid, Spain}

\author{I. Calvo}
\affiliation{Laboratorio Nacional de Fusi\'on, CIEMAT, 28040 Madrid, Spain}

\author{F.I. Parra}
\affiliation{Princeton Plasma Physics Laboratory, Princeton, NJ 08540, USA}

\author{F.J. Escoto}
\affiliation{Laboratorio Nacional de Fusi\'on, CIEMAT, 28040 Madrid, Spain}

\author{E. S\'anchez}
\affiliation{Laboratorio Nacional de Fusi\'on, CIEMAT, 28040 Madrid, Spain}

\author{H. Thienpondt}
\affiliation{Laboratorio Nacional de Fusi\'on, CIEMAT, 28040 Madrid, Spain}

\date{\today}

\maketitle

This Letter has presented piecewise omnigenous (pwO) fields, a new family of optimized stellarator fields with levels of collisional energy transport similar to that of tokamaks and omnigenous stellarators, while not being subject to some of the constraints that omnigenity imposes on the spatial variation of the magnetic field strength $B$ [1]. This result may radically broaden the space of accessible reactor-relevant configurations. Naturally, the next step is a rigorous assessment of other physics properties of nearly pwO configurations, in order to qualify them as serious contenders to quasi-isodynamic or quasisymmetric configurations.

In the next paragraphs we provide preliminary indications, based on state-of-the-art published studies and new additional simulations, that there exist nearly pwO configurations that meet reactor design criteria other than low neoclassical transport. We will mainly refer to two magnetic configurations: LHD inward-shifted and configuration A of~[16]. Both are relatively close to being pwO (see figure 5 of the Letter) and have consequently low neoclassical transport, compatible with a reactor scenario (in fact, LHD-like devices have been thoroughly studied as reactor candidates, both from the physics [30] and engineering \cite{sagara2008reactor} points of view). The \textit{standard} configuration of LHD, which is further from being pwO (see figure 2 in [27]) will be used in some cases as a reference, as it differs from the inward-shifted configuration in its neoclassical optimization, but shares other properties, such as the aspect ratio or the rotational transform profile.

\setcounter{figure}{5}
\appendix

\vspace{-.6cm}

\section{Coils}\label{SEC_COILS}

In a stellarator configuration that is very close to exact piecewise omnigenity, the magnetic field strength has to vary on a very small length scale, which could in principle lead to complicated magnets. Nevertheless, this is not the case for the LHD inward-shifted configuration. In the experiment, its field is mainly produced by superconducting helical coils. This concept presents advantages with respect to modular coils in terms of shaping, plasma accessibility and room for a blanket \cite{sagara2008reactor}; their length is considered a drawback, but this could be addressed in a reactor by means of segment fabrication of high-temperature superconducting coils with low-resistance joints \cite{ito2021segments}. Recently, several configurations of LHD have been numerically produced also with modular coils in~\cite{kappel2024gradient}. More importantly, no correlation was observed in the configuration space of LHD between the level of neoclassical optimization and the difficulty in obtaining a set of coils.

\begin{figure}
\includegraphics[angle=0,width=0.5\columnwidth]{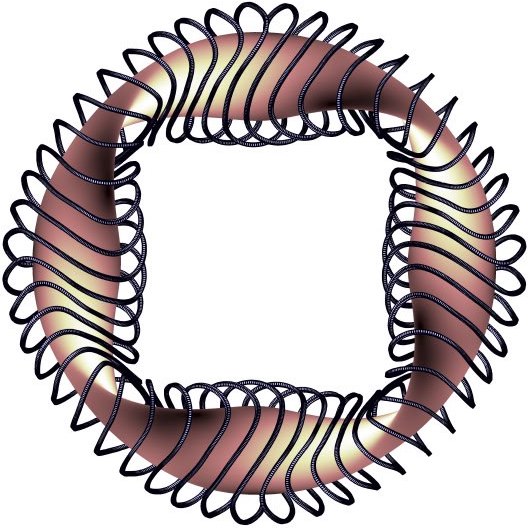}
\caption{Last closed flux surface of configuration A with filamentary coils. \label{FIG_COILS}}
\end{figure}

Even closer to piecewise omnigenity, figure~\ref{FIG_COILS} shows a set of modular coils (6 per half-period) created with the code~\texttt{FOCUS}~\cite{zhu2017focus}. They reproduce the field of configuration A (scaled to a reactor with minor radius $a=1.7\,$m) with a surface-averaged normalized field error of 3.5$\times 10^{-3}$, thus sufficiently maintaining its good confinement properties (see below). The engineering parameters (the minimum coil-coil distance is 72$\,$cm, and the minimum coil-plasma distance is 1.4$\,$m) are in principle compatible with a reactor design. Even improved parameters could likely be achieved with a thorough optimization of the coils, which is out of the scope of this work.

\vspace{-.3cm}

\section{Energetic ions}\label{SEC_FI}

\begin{figure}[ht!]
\includegraphics[angle=0,width=0.85\columnwidth]{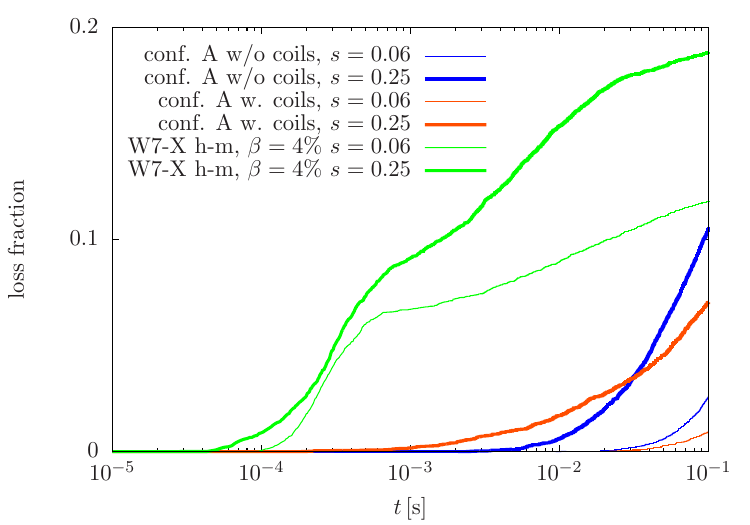}
\caption{Energetic ion losses for configuration A.\label{FIG_FI}}
\end{figure}

Reduced neoclassical transport of energy does not guarantee good confinement of the energetic ions, see e.g. [13], as their neoclassical transport is not only determined by the (orbit-averaged) radial component of magnetic drift, but also by the component that is tangent to the flux-surfaces. The latter is connected to the radial variation of $B$, see e.g.~\cite{velasco2021prompt}. Both configuration A and LHD inward-shifted present good energetic ion confinement. In fact, the coils of figure~\ref{FIG_COILS} are able to reasonably maintain the good energetic ion confinement of the fixed-boundary magnetohydrodinamic equilibrium presented in [16]. This is shown in figure~\ref{FIG_FI} by means of new collisionless guiding-center simulations of alpha-particles with \texttt{ASCOT} (launched at two different flux surfaces with a uniform angular distribution and an isotropic distribution in velocities).

The time dependence of the energetic ion losses is remarkable (compare with W7-X high-mirror, reactor-scaled, in the same figure), and it is likely a general property of nearly pwO configurations, as it is also observed in LHD inward-shifted [31]. A theory-based explanation can be advanced: the smallness of the orbit-averaged radial magnetic drift (in combination with some radial variation of $B$) in regions of the flux-surface far from transitions contributes to prevent prompt losses ($t\lesssim 10^{-3}\,$s). At longer times, the accumulation of small radial excursions around  transitions (in a field like the one of figure 3 top right, in which the $B$-contours do not have corners, but are slightly curved) could drive losses of partially thermalized energetic ions via stochastic diffusion~\cite{beidler2001stochastic}. In a reactor based on a nearly pwO configuration, this qualitative behaviour should limit the damage to plasma-facing components and could mitigate the retention of helium ashes without a strong impact on alpha heating efficiency.

\vspace{-.3cm}

\section{Turbulence}\label{SEC_TURB}

A direct connection between turbulence and the on-surface variation of $B$ of a given magnetic configuration cannot be drawn. For instance, the heat flux of two configurations whose $B$ has toroidal symmetry, NCSX and ASDEX Upgrade, present a very different dependence on the plasma gradients~\cite{thienpondt2024comp}. Moreover, stellarators may have very different stability properties depending on the radial variation of $B$~\cite{rosenbluth1968maxJ,helander2013tem}, of the rotational transform (i.e., the magnetic shear)  or of the radial electric field.

Cross-device systematic gyrokinetic studies, e.g.~\cite{thienpondt2024comp}, have only recently become computationally possible. In agreement with previous work (see below), the results of~\cite{thienpondt2024comp} confirms that LHD has relatively low levels of ion-temperature-gradient (ITG) turbulence, a transport channel that has limited the performance of W7-X~[33]. 

\begin{figure}[ht!]
\includegraphics[angle=0,width=0.85\columnwidth]{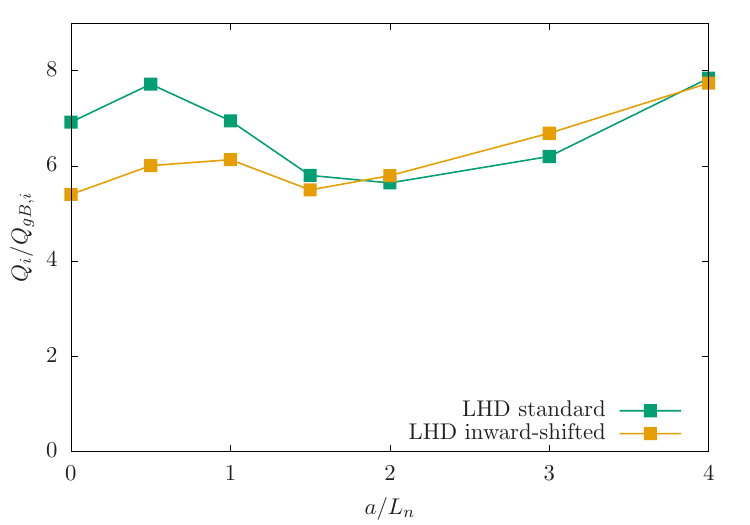}
\caption{Normalized turbulent ion heat flux in LHD.\label{FIG_GK}}
\end{figure}

In figure \ref{FIG_GK}, we extend the work of~\cite{thienpondt2024comp} with additional electrostatic non-linear gyrokinetic simulations with the code \texttt{stella} (the numerical setup is detailed in~\cite{thienpondt2024comp}) in the configuration space of LHD. The simulations include kinetic electrons, and the electron and ion normalized temperature gradients are $a/L_{T_i}=a/L_{T_e}=3$. For small normalized density gradients $a/L_{n}$, the  inward-shifted configuration displays a smaller heat flux. This is in line with previous works that showed reduced ITG transport in the neoclassically-optimized configuration, and with experimental observations  (see e.g.~\cite{sugama2008is} and \cite{motojima2003inward}, respectively). Little dependence on the closeness to piecewise omnigenity is observed for large $a/L_{n}$. 

In summary, while  a general prediction of the turbulent transport properties of the family of nearly pwO configurations is extremely complicated, the results presented in figure \ref{FIG_GK} provide a preliminary indication that some nearly pwO configurations could exhibit relatively low levels of turbulent transport, at least in some parameter regimes.

\vspace{-.8cm}

\bibliography{/Users/velasco/Work/PAPERS/bibliography.bib}